\documentclass[proof]{pasj01}
\draft 
\Received{$\langle$reception date$\rangle$}
\Accepted{$\langle$acception date$\rangle$}
\Published{$\langle$publication date$\rangle$}

\begin{document}

\title{High-resolution spectroscopy of the extremely iron-poor post-AGB star CC Lyr}
\author{Wako \textsc{Aoki}\altaffilmark{1}\altaffilmark{2}}
\altaffiltext{1}{National Astronomical Observatory, 
 2-21-1 Osawa, Mitaka, Tokyo 181-8588, Japan }
\email{aoki.wako@nao.ac.jp}

\author{Tadafumi \textsc{Matsuno}\altaffilmark{2}}
\altaffiltext{2}{Department of Astronomical Science, School of Physical Sciences, The Graduate University of Advanced Studies (SOKENDAI), 2-21-1 Osawa, Mitaka,
Tokyo 181-8588, Japan}
\email{tadafumi.matsuno@nao.ac.jp}

\author{Satoshi \textsc{Honda}\altaffilmark{3}}
\altaffiltext{3}{Center for Astronomy, University of Hyogo, 407-2, Nishigaichi, Sayo-cho, Sayo, Hyogo 679-5313, Japan}
\email{honda@nhao.jp}

\author{Mudumba \textsc{Parthasarathy}\altaffilmark{4}}
\altaffiltext{4}{Indian Institute of Astrophysics, II Block, Koramangala, Bangalore 560 034, INDIA}
\email{m-partha@hotmail.com}

\author{Haining \textsc{Li}\altaffilmark{5}}
\altaffiltext{5}{Key Lab of Optical Astronomy, National Astronomical
  Observatories, Chinese Academy of Sciences, A20 Datun Road,
  Chaoyang, Beijing 100012, China}
\email{lhn@nao.cas.cn}

\author{Takuma \textsc{Suda}\altaffilmark{6}}
\altaffiltext{6}{Research Center for the Early Universe, University of Tokyo, 7-3-1 Hongo, Bunkyo-ku, Tokyo 113-0033, Japan}
\email{suda@resceu.s.u-tokyo.ac.jp}



\maketitle

\begin{abstract}
High-resolution optical spectroscopy was conducted for the metal-poor
post-AGB star CC Lyr to determine its chemical abundances and spectral
line profiles. Our standard abundance analysis confirms its extremely
low metallicity ([Fe/H]$<-3.5$) and a clear correlation between
abundance ratios and the condensation temperature for 11 elements,
indicating that dust depletion is the cause of the abundance anomaly
of this object.  The very low abundances of Sr and Ba, which are
detected for the first time for this object, suggest that heavy
neutron-capture elements are not significantly enhanced in this object by the
s-process during its evolution through AGB phase. Radial velocity of
this object and profiles of some atomic absorption lines show 
variations depending on pulsation phases, which could be formed by
dynamics of the atmosphere rather than by binarity or contributions
of circumstellar absorption.  On the other hand, the H~$\alpha$
emission with double peaks shows no evident velocity shift, suggesting
that the emission is originating from the circumstellar matter, presumably
the rotating disk around the object.
\end{abstract}

\section{Introduction}

A small number of warm post-AGB stars are known to show anomalously
low abundances of Fe and other metals. BD+39$^{\circ}$4926, HR~4049,
HD~44179 and HD~52961 are known to have [Fe/H]$<-3$
(\cite{waelkens91,vanwinckel95}). Most metals including Fe are highly
depleted in these stars, whereas some volatile elements, e,g., C, O,
S, Zn, are not depleted as significant as Fe, indicating that formation
and ejection of dust from these objects are the cause of depletion of
refractory elements. 

Although no convincing models have yet been established, scenarios to
explain this phenomenon have been proposed. The basic idea is that,
after a circumstellar shell is formed by efficient mass-loss during
the evolution along the asymptotic giant branch, the gas component is
re-accreted onto the surface of the star. The dust component that
contains most metals of the ejected material is removed in the process
(\cite{mathis92}). This specific process may operate in circumstellar
disk that is formed in a binary system with a low-mass companion
star (\cite{waters92}).  Indeed, binarity is confirmed for the four
extremely metal-poor post-AGB stars mentioned above
(\cite{vanwinckel95}).

CC Lyr is a high galactic latitude star  ($b=17.23^{\circ}$) and is classified as a Type II Cepheid based on its light curve
(e.g., \cite{schmidt04}). The abundance analysis for this object by
\citet{maas07}, however, revealed that refractory elements show large
depletion (e.g., [Fe/H]$=-3.5$), as found for some post-AGB stars as
mentioned above, indicating that this object is an extremely
metal-poor post-AGB star.

Other features of this object also support the interpretation that
this object is a post-AGB star. One is the pulsation and its period
(24.23~days), which is similar to the periods previously found for
other metal-poor post-AGB stars. A significant excess of infrared
emission is also found, as the case of other post-AGB stars. We note
that the class of Type II Cepheides consists of many types of objects,
including post-AGB stars like RV Tau stars (\cite{wallerstein02}).


This object was re-discovered as a candidate of extremely metal-poor
star by the spectroscopic survey with the Large sky Area Multi-Object
fiber Spectroscopic Telescope (LAMOST; \cite{cui12}). High-resolution
spectra of this object were obtained with the Subaru Telescope during
a follow-up program. We here report our new measurement of chemical
abundances, radial velocities and absorption line profiles of this
object.

\section{Observations}

The object was identified to be a candidate of a metal-poor star by
the medium resolution ($R\sim 1800$) spectroscopy with LAMOST in its
regular survey (\cite{zhao12}). The ID in the LAMOST survey is LAMOST
J~1833+3138.  Figure~1 shows the LAMOST spectrum of this object. The
spectral features are very similar to those of carbon-enhanced
metal-poor (CEMP) stars, except for the emission feature of H$\alpha$.

High resolution ($R=45,000$) spectra of this object were obtained with
the Subaru Telescope High Dispersion Spectrograph (HDS;
\cite{noguchi02}) in a program (S16A-119I) for follow-up spectroscopy
for metal-poor star candidates found with LAMOST (\cite{li15}). Since
the object is bright ($V\sim 12$), observations were conducted using
the twilight time in the Subaru program.

Observations with the Subaru/HDS were made at three epochs in April
and May 2016 (Table~\ref{tab:obs}). The pulsation phase is calculated
using the light curve parameters obtained by \citet{schmidt04}. Recent
photometry data provided by AAVSO \footnote{https://www.aavso.org}
suggest that the phase possibly has offset by at most 0.1, but
confirm that our observations were made in the late half of the phase,
and the first two were done at almost the same phase.

Standard data reduction procedures were carried out with the IRAF
echelle package\footnote{IRAF is distributed by the National Optical
  Astronomy Observatories, which is operated by the Association of
  Universities for Research in Astronomy, Inc. under cooperative
  agreement with the National Science Foundation.}. The wavelength
shift due to earth's orbital motion is corrected using the IRAF task {\it
  rvcor}.

\section{Chemical composition}\label{sec:cc}

\subsection{Measurements of spectral lines}

The absorption lines measured from the spectra are given in
Table~\ref{tab:ew}. Many C {\small I} lines are detected in the
spectra, as expected from the previous studies. Line data of most of
the C {\small I} lines are taken from \citet{takeda02} who have studied
the spectrum of the metal-poor post-AGB star HR~4049. Line data for
other spectral features are taken from studies on very metal-poor
stars (e.g., \cite{aoki13}).

Strong absorption features have asymmetry as shown in
Figure~\ref{fig:c1}. The asymmetry is in particular evident in the first two
spectra obtained at $\phi\sim 0.7$, whereas the lines are almost
symmetrical in the other spectrum. The variation of the line profiles is
discussed in \S~\ref{sec:dyn}.

We measure equivalent widths ($W$) by fitting a Gaussian profile.
Since strong lines in the spectrum of April 27 show asymmetry as depicted
in Figure~\ref{fig:c1}, we measure the equivalent widths by direct
integration for lines with $\log W/\lambda >-5.0$ in the spectrum.
The measured equivalent widths are given in Table~\ref{tab:ew} with
line data used in the abundance analysis.

\subsection{Stellar parameters and abundance analysis}

Model atmospheres of the ATLAS NEWODF grid (\cite{castelli97}) are
applied to our 1D/LTE analysis. 

We started the analysis of the spectrum obtained on 27 April adopting
the stellar parameters given by \citet{maas07}: $T_{\rm eff}=6250$~K,
$\log g=1.0$ and $v_{\rm turb}$=3.5~km~s$^{-1}$. We found that the Fe
abundances derived from the Fe {\small I} lines is about 0.5~dex higher than
that from the Fe {\small II} ones for this parameter set. This discrepancy is
reduced by assuming higher gravity and lower effective
temperature. Hence, we adopt $T_{\rm eff}=6000$~K and $\log
g=2.0$. Abundance results are given in Table~\ref{tab:abundance}.

As found by \citet{maas07}, this object has CNO abundances as high as
that of the Sun, but very low abundances of most of other metals including
Fe. Hence, its chemical composition is not approximated by the
scaled-solar metal abundances. We examined model atmospheres for two
extreme cases of metallicity, [M/H]$=-0.5$ and $-3.5$. The difference
of elemental abundances obtained by the two models is less than
0.2~dex, and less than 0.1~dex in most cases (Table~\ref{tab:error}),
which does not affect the following discussions. We adopt the model of
[X/H]$=-0.5$ to obtain the final result.

The microturbulent velocity is uncertain due to the limited number of
spectral lines used for deriving abundances for each element. We
modified the value as the derived carbon abundances from individual C
{\small I} lines is independent of the strengths of the lines used in
the analysis. The adopted value is $v_{\rm turb}$=4.5~km~s$^{-1}$,
which is larger than that of \citet{maas07}. The effect of the changes
of $v_{\rm turb}$ on the derived abundances is, however, almost
negligible (Table~\ref{tab:error}).
  


 

The carbon abundance is determined by the measurements of 13 C {\small
  I} lines. The abundance is also estimated by the comparison of
synthetic spectra for CH molecular band at 4322~{\AA} with observed
ones (Figure~\ref{fig:ch}). The carbon abundance assumed for the
calculation for the spectrum obtained on April 27 is [C/H]$=-0.33$,
which agrees with the result from C {\small I} lines within the
measurement error. The CH band is sensitive to the effective
temperature adopted by the analysis. The agreement of the abundances
from C {\small I} lines and the CH band supports the effective
temperature adopted in the analysis.


The abundances of other elements are derived by the standard analysis
for the measured equivalent widths. 

The spectrum obtained on May 21 is almost identical to that of April
27, and the S/N ratio is not as good as the first one. We do not
conduct abundance analysis for the May 21st spectrum. On the other hand, the
spectrum of May 29 shows significant changes of absorption features
from the other two spectra. As found in Figure~\ref{fig:ch}, the CH
band is stronger in this spectrum, indicating that the temperature is
lower at this epoch. We assume $T_{\rm eff}=5750$~K for the analysis
of the May 29 spectrum.

The equivalent widths measured for lines with high excitation
potential in the May 29 spectrum are smaller than those in the April
27th one, whereas the trend is opposite for lines with low excitation
potential (Table~\ref{tab:error}). This support the lower $T_{\rm eff}$ on May 29. Thanks to
the lower temperature, the two Ba {\small II} lines are detected in
the spectrum of May 29. The two Sr {\small II} resonance lines are
clearly detected in both spectra. The abundances of heavy
neutron-capture elements are measured for the first time for CC Lyr.
The effect of hyperfine splitting is ignored because the spectral
lines analyzed in the present work are sufficiently weak.

\subsection{Abundance ratios and correlation with condensation temperature}

The elemental abundances derived by our analysis are plotted as a
function of the condensation temperature in Figure~\ref{fig:abtc}. As
found by \citet{maas07}, a clear correlation between the abundances
and the condensation temperature is found. This correlation confirms
that the refractory elements with high condensation temperature are
severely depleted by condensation to dust grains that are removed from
the material preserved at the surface of the star.

The S and Zn abundances are slightly lower than the solar value, and
the ratio [Zn/S] is about $-0.5$. These values are along the trend
found by previous studies for metal-poor post-AGB stars including CC
Lyr (\cite{takeda02, maas07}). If the abundances of these two elements
with low condensation temperature primarily trace the initial
metallicity of this object, it should be a slightly metal-poor, and
would be a low-mass object with a long life-time. This is consistent
with the masses of metal-poor post-AGB stars estimated by
\citet{bono97} and \citet{gingold85} to be in the range of 0.52 to 0.59
M$_{\odot}$.


Our measurements provide new data points of heavy neutron-capture
elements, Sr and Ba, which could be produced by the s-process during
the evolution along the asymptotic giant branch. There is no excess of
these elements in this object compared to the trend in
Figure~\ref{fig:abtc}: if the correlation between the condensation
temperature and elemental abundances found for most metals are
extended to those of Sr and Ba, which have the highest condensation
temperature, additional s-process contribution to this object should
be small if any. The typical mass of AGB stars in which s-process is
active is estimated to be 1.5--3~M$_{\odot}$ (\cite{kappeler11}). 
The minimum core mass to trigger the third dredge-up predicted by
stellar evolutionary models ranges between 0.5-0.6 M$_{\odot}$
(\cite{stancliffe05}; \cite{weiss09}; \cite{cristallo11}) or higher
(\cite{iben81}). This is also consistent with the mass estimates for
metal-poor post-AGB stars mentioned above. 



\section{Dynamics}\label{sec:dyn}

In the high resolution spectra of CC Lyr obtained by the present work,
variations and asymmetry of absorption features are found. H$\alpha$
is observed as emission with double peaks. This section reports these
spectral features for the spectra obtained at the three epochs. We
note that no clear circumstellar absorption of C$_{2}$ molecule is
detected for CC Lyr, in contrast to another very metal-poor AGB star
HD~52961 (\cite{bakker96}). Na {\small I} D lines show complicated
absorption features, in addition to the photospheric absorption. The
absorption features, however, appear at longer wavelengths than the
photospheric ones, which is not simply explained by circumstellar
absorption formed in an expanding shell. These absorption features
would be primarily due to interstellar absorption, as found in the Na
{\small I} of HR~4049 (\cite{bakker98}).

\subsection{Radial velocities}

The radial velocity of the object is measured by the wavelength shift
of the core of the six absorption lines of Mg, Fe and Zn. The derived
values are given in Table~\ref{tab:obs}. The error given in the table
is standard deviation of the measurements. Including possible
systematic errors due to instability of the spectrograph, the
uncertainty of the measurement could be as large as
0.5~km~s$^{-1}$. 


The three epochs of the observations correspond to the phase of the
light curve ($\phi$) of 0.7, 0.7 and 1.0 for April 27, May 21, and May
29, respectively. The radial velocities measured for the first two
observations at $\phi=0.7$ are almost identical, whereas the other at
$\phi=1.0$ is 11~km~s$^{-1}$ larger. This suggests that the radial
velocity change is related to the pulsation of the object, although
definitive conclusion cannot be derived from the three measurements.

Taking the binarity found for other metal-poor post-AGB stars into
consideration, this object is expected to belong to a binary system
with an orbital period of several hundred days
(\cite{vanwinckel95}). In order to examine the binarity of this object
from radial velocity changes, long-range monitoring that can
distinguish the effect of the pulsation and orbital motion in the
binary system is required.

\subsection{Asymmetry of absorption profiles}

Strong C {\small I} lines show asymmetry in the spectra obtained on
April 27 and May 21. Figure~\ref{fig:c1} shows the line profiles in a
velocity scale. The profile is empirically reproduced by assuming two
components of Gaussian with 15~km~s$^{-1}$ separation. 
Similar asymmetry of absorption line profile is found for
HD~52961. \citet{kipper13} analyzed the profile assuming two Gaussian
components and discussed the weaker blue component as absorption by
expanding circumstellar shell. 

The asymmetry is, however, not evident in the spectrum of CC Lyr
obtained on May 29 at a different pulsation phase
(Figure~\ref{fig:c1}). There is no detectable absorption at the
velocity of $\sim -50$~km~s$^{-1}$ at which a ``blue component'' is
found in the other two spectra. This indicates that the blue component
is also formed in the dynamic photosphere, or the expanding
circumstellar shell appears only in a limited phases of the pulsation.

\subsection{H$\alpha$ emission}

Figure~\ref{fig:ha} shows the velocity profile of the H$\alpha$
emission in the spectra. By contrast to the C {\small I} absorption,
H$\alpha$ emission shows no detectable velocity shift, although the
profile and strength of the double peaks slightly change.


CC Lyr seems to be very similar to the very metal-poor post-AGB binary
stars HD 52961 and IRAS 11472-0800
([Fe/H]$=-2.7$; \cite{vanwinckel12}). The H$\alpha$ profile of IRAS
11472-0800 with double peaks and its variation reported by
\citet{vanwinckel12} seems to be very similar to those of CC Lyr.


Emission of H$\alpha$ is also observed in the very metal-poor post-AGB
star HR 4049 (\cite{bakker98}). The feature found for HR~4049 consists
of strong emission of the red component, and the weak blue component
shows significant variation, which seems to be more complicated than
the emission feature of CC Lyr. \citet{bakker98} discuss two possible
models to explain the H$\alpha$ emission of HR~4049: one is the
circumbinary disk that reflects the light of the post-AGB star. The
variation of the emission feature is explained by assuming that only a
small portion of the disk reflects the light. If similar model is
applied to the H$\alpha$ emission of CC Lyr, the clear double peak,
without significant time variation, suggests that a larger part of the
disk reflects the star light. The other model is the activity in
the extended atmosphere like the chromosphere. In the case of CC Lyr, no
correlation between H$\alpha$ and the pulsation period suggests this
model is unlikely, though further monitoring of the emission feature is
required to clearly distinguish the two models.


\section{Discussion and concluding remarks}

Our abundance measurements for CC Lyr based on high resolution
spectroscopy confirmed the low abundances of metals with high
condensation temperatures, that indicate significant effects of dust
depletion on the surface composition of this object, as found in
several post-AGB stars with extremely low metallicity
(\cite{vanwinckel95}). Low abundances of Sr and Ba, which are
  detected for the first time for this object in the phase with low
  temperature suggests that s-process was not very effective during its
  AGB phase.

CC Lyr is a very metal-poor post-AGB supergiant similar to HD 52961
and IRAS~11472-0800, which have relatively low temperatures ($T_{\rm
  eff}\sim 6000$~K) among several objects in this class. The radial
velocity variations indicate that it may be a single lined
spectroscopic binary similar to HD 52961 and other very metal-poor
post-AGB stars. However, to disentangle the radial velocity variations
due to pulsation and binarity needs very long period monitoring
of the radial velocity variations. 


Our high-resolution spectroscopy for CC Lyr also revealed the
asymmetry of absorption features and their variation with pulsation
phases. This is similar to that found in the very metal-poor post-AGB
star HD~52961. Similar line profile asymmetry seems to be common in
post-AGB stars, as they have very extended atmospheres and velocity
fields due to mass motions etc. Plane parallel approximation and LTE
conditions may not be very suitable to analyze the spectra of these
stars. However, the agreements of abundance results for different
pulsation phases support that the overall abundance features found in
the present work, in particular the correlation with condensation
temperatures of elements, are robust.

The emission feature of H$\alpha$ found in CC Lyr could be key to
probing the structure of circumstellar disk. The double peak emission
does not show clear correlation with the pulsation phase, although the
number of observations is still limited. The feature is similar to
those found in other post-AGB stars, in particular IRAS~11472-0800
(\cite{vanwinckel12}). Monitoring H$\alpha$ emission for CC Lyr and
other metal-poor post-AGB stars, as done by \citet{bakker98} for
HR~4049, will help us understand the circumstellar matter of post-AGB
stars in general.

The group of post-AGB stars with very low metallicity still consists
of a small number of objects including CC Lyr. Their features are,
however, more or less found in many post-AGB stars.  The pulsation
period of CC Lyr (24.3 days) and amplitude of the light curve of CC
Lyr indicates it is in the border region of W Vir type Type II
Cepheids and RV Tau stars (\cite{wallerstein02}).  H$\alpha$ emission
profile is observed in the spectra of many W Vir type Type II
Cepheids.  \citet{lemasle15} find H$\alpha$ emission profile somewhat
similar to that found in the spectrum of CC Lyr in a few Type II
Cepheids of W Vir class. They also find mild depletion of refractory
elements indicating dust gas separation. \citet{gonzalez96} found that
the Type II Cepheid ST Pup is a single lined spectroscopic binary with
a dusty disk and depletion of Fe and other refractory elements similar
to that of CC Lyr. In more general, interaction of circumstellar
matter and a companion star in a binary system seems to play essential
role in the evolution of post-AGB star and formation of planetary
nebulae. Continuous study of metal-poor post-AGB stars like CC Lyr
will contribute to the understanding of the late phase of low-mass
star evolution.

\begin{ack}
This work is based on data collected at the Subaru Telescope, which is
operated by the National Astronomical Observatory of Japan. 
WA and TS were supported by JSPS KAKENHI Grant Numbers JP16H02168.
MP was supported by the NAOJ Visiting Fellow Program of the Research Coordination Committee, National Astronomical Observatory of Japan (NAOJ), National Institutes of Natural Sciences(NINS).
\end{ack}

\clearpage

\begin{table}
\tbl{Observations 
\label{tab:obs}}{%
\begin{tabular}{@{}lc@{\qquad}lc@{}}
\hline\noalign{\vskip 3pt}
\multicolumn{1}{c}{Observation date (UT)}    &  \multicolumn{1}{c}{HJD} &  \multicolumn{1}{c}{Phase} &  \multicolumn{1}{c}{$V_{\rm Helio}$(km~s$^{-1}$)} \\
\hline\noalign{\vskip3pt} 
April 27, 2016 & 2457506.09 & 0.68 & $-36.5\pm 0.2$ \\
May 21, 2016 & 2457530.00 & 0.67 & $-36.1\pm 0.6$ \\
May 29, 2016 & 2457538.00 & 1.00 (0.00) & $-25.1\pm 0.3$ \\
\hline\noalign{\vskip 3pt} 
\end{tabular}
}
\end{table}

\begin{longtable}{lccccc}
  \caption{Equivalent widths}\label{tab:ew}
  \hline              
 Ion  & Wavelength ({\AA}) & $\log gf$ & $\chi$(eV) & $W$(m{\AA})April 27 & $W$(m{\AA})May 29 \\
\endhead
  \hline
\endfoot
  \hline
\endlastfoot
  \hline
    C I &  4766.67 & $-$2.51 &  7.48 & ..... & 16.3 \\
    C I &  4770.03 & $-$2.33 &  7.48 & 30.7 & 26.3 \\
    C I &  4771.74 & $-$1.76 &  7.49 & 96.1 & 64.7 \\
    C I &  4775.90 & $-$2.19 &  7.49 & 48.3 & 33.7 \\
    C I &  4817.37 & $-$2.89 &  7.48 & 19.0 & 11.7 \\
    C I &  4826.80 & $-$3.05 &  7.49 & ..... &  9.8 \\
    C I &  4932.05 & $-$1.66 &  7.69 & 73.1 & 52.9 \\
    C I &  5023.85 & $-$2.21 &  7.95 & 13.5 & ..... \\
    C I &  5024.92 & $-$2.73 &  7.95 &  4.3 & ..... \\
    C I &  5039.06 & $-$1.79 &  7.95 & 36.7 & ..... \\
    C I &  5040.13 & $-$2.30 &  7.95 & 20.5 & ..... \\
    C I&   5052.17 & $-$1.30 &  7.69 & 129.4 & 79.5 \\
    C I &  5793.12 & $-$2.16 &  7.95 & 20.4 & ..... \\
    C I &  6013.21 & $-$1.47 &  8.65 & ..... & 13.7 \\
    C I &  6014.83 & $-$1.71 &  8.64 & 17.9 & ..... \\
    C I &  6587.61 & $-$1.60 &  8.54 & 48.1 & ..... \\
    O I &  5577.34 & $-$8.20 &  1.97 & 20.8 & 26.1 \\
    O I &  6300.30 & $-$9.82 &  0.00 & 72.6 & 78.0 \\
    O I &  6363.78 & $-$10.30 & 0.02 & 19.6 & 35.4 \\
   Na I &  5682.63 & $-$0.70 &  2.10 & 11.1 & 21.0 \\
   Na I &  5688.00 & $-$0.46 &  2.10 & 18.2 & ..... \\
   Mg I &  5167.32 & $-$0.86 &  2.71 & 22.0 & 42.8 \\
   Mg I &  5172.68 & $-$0.45 &  2.71 & 45.6 & 59.1 \\
   Mg I &  5183.60 & $-$0.24 &  2.72 & 68.6 & 72.0 \\
    S I &  4694.11 & $-$1.77 &  6.52 & 21.9 & 13.3 \\
    S I &  4695.44 & $-$1.92 &  6.52 & 11.3 &  9.5 \\
    S I &  4696.25 & $-$2.14 &  6.52 &  5.9 & ..... \\
    S I &  6052.67 & $-$0.74 &  7.87 & 18.4 & 16.8 \\
    S I &  6757.17 & $-$0.31 &  7.87 & 43.4 & 33.9 \\
   Mn I &  4030.75 & $-$0.50 &  0.00 & 23.0 & 50.7 \\
   Mn I &  4033.06 & $-$0.62 &  0.00 & 18.6 & 37.0 \\
   Fe I &  4383.54 &  0.21 &  1.49 & 54.7 & 74.3 \\
   Fe I &  4404.75 & $-$0.15 &  1.56 & 30.0 & 47.1 \\
   Fe I &  4415.12 & $-$0.62 &  1.61 & 19.0 & 28.3 \\
   Fe I &  4957.60 &  0.23 &  2.81 & 11.5 & ..... \\
   Zn I &  4680.13 & $-$0.81 &  4.01 & 17.9 & 32.0 \\
   Zn I &  4722.15 & $-$0.39 &  4.03 & 47.3 & 61.3 \\
   Zn I &  4810.53 & $-$0.17 &  4.08 & 60.2 & 78.2 \\
   Zn I &  6362.34 &  0.15 &  5.80 & ..... & 14.8 \\
  Fe II &  4923.93 & $-$1.21 &  2.89 & 20.5 & 49.0 \\
  Fe II &  5169.00 & $-$0.87 &  2.89 & 40.6 & 21.6 \\
  Sr II &  4077.71 &  0.16 &  0.00 & 45.7 & 62.9 \\
  Sr II &  4215.52 & $-$0.16 &  0.00 & 22.1 & 49.2 \\
  Ba II &  4554.03 &  0.16 &  0.00 & ..... &  8.8 \\
  Ba II &  4934.08 & $-$0.15 &  0.00 & ..... &  7.1 \\
\end{longtable}

\begin{table}
\tbl{Chemical Composition of CC Lyr
\label{tab:abundance}}{%
\begin{tabular}{llcccccccc}
\hline\noalign{\vskip 3pt}
& & & \multicolumn{3}{c}{April 27 data} &  &  \multicolumn{3}{c}{May 29 data} \\
\hline\noalign{\vskip 3pt} 
$T_{\rm eff}$ & & & \multicolumn{3}{c}{6000~K} && \multicolumn{3}{c}{5750~K} \\
$\log g$  & & & \multicolumn{3}{c}{2.0} && \multicolumn{3}{c}{2.0} \\
{[X/H]}  & & & \multicolumn{3}{c}{$-0.5$} && \multicolumn{3}{c}{$-0.5$} \\
$v_{\rm turb}$  & & & \multicolumn{3}{c}{4.5~km~s$^{-1}$} && \multicolumn{3}{c}{4.5~km~s$^{-1}$} \\
\cline{4-6}\cline{8-10}
Elemet & Ion & $\log\epsilon_{\odot}$ & [X/H] & $n$  & $\sigma_{\rm total}$ &  &  [X/H] & $n$  & $\sigma_{\rm total}$  \\
\hline\noalign{\vskip 3pt} 
C  & C I   & 8.43 & $-$0.11 & 13 & 0.18 &  & $-$0.23 & 9 & 0.20 \\
O  & O I   & 8.69 & 0.35  & 3  & 0.23 &  & 0.40  & 3 & 0.23 \\
Na & Na I  & 6.24 & $-$1.35 & 2  & 0.18 &  & $-$1.13 & 1 & 0.18 \\
Mg & Mg I  & 7.60 & $-$3.64 & 3  & 0.22 &  & $-$3.60 & 3 & 0.24 \\
S  & S I   & 7.12 & $-$0.39 & 5  & 0.12 &  & $-$0.42 & 4 & 0.15 \\
Mn & Mn I  & 5.53 & $-$3.54 & 2  & 0.33 &  & $-$3.40 & 2 & 0.36 \\
Fe & Fe I  & 7.50 & $-$3.85 & 4  & 0.26 &  & $-$3.90 & 3 & 0.31 \\
Fe & Fe II & 7.50 & $-$4.04 & 2  & 0.21 &  & $-$4.09 & 2 & 0.22 \\
Zn & Zn I  & 4.56 & $-$1.02 & 3  & 0.22 &  & $-$0.89 & 4 & 0.21 \\
Sr & Sr II & 2.87 & $-$4.35 & 2  & 0.26 &  & $-$4.17 & 2 & 0.27 \\
Ba & Ba II & 2.18 & .....   & ... & ..... &  & $-$4.22 & 2 & 0.25 \\ 
\hline\noalign{\vskip 3pt} 
\end{tabular}
}
\end{table}

\begin{table}
\tbl{Abundance sensitivity to stellar parameters 
\label{tab:error}}{%
\begin{tabular}{lccccc}
\hline\noalign{\vskip 3pt}
 Element  & $\delta T_{\rm eff}$ & $\delta\log g$ &  $\delta v_{\rm turb}$ & $\delta$[M/H] &  r.s.s. \\
   & $+250$~K & $-0.5$~dex & $-0.5$~km~s$^{-1}$ & $-3.0$~dex & \\
\hline\noalign{\vskip 3pt}
C  & $-$0.08 & $-$0.14 & 0.02 &  0.07 & 0.17 \\
O  &  0.11 & $-$0.16 & 0.01 & $-$0.05 & 0.20 \\
Na &  0.10 &  0.01 & 0.00 & $-$0.08 & 0.13 \\
Mg &  0.15 &  0.00 & 0.01 & $-$0.12 & 0.20 \\
S  & $-$0.02 & $-$0.10 & 0.01 &  0.02 & 0.10 \\
Mn &  0.24 &  0.00 & 0.01 & $-$0.19 & 0.30 \\
Fe &  0.19 &  0.00 & 0.01 & $-$0.16 & 0.25 \\
Fe &  0.04 & $-$0.17 & 0.01 & $-$0.04 & 0.17 \\
Zn &  0.15 & $-$0.01 & 0.01 & $-$0.12 & 0.20 \\
Sr &  0.13 & $-$0.17 & 0.01 & $-$0.09 & 0.23 \\
Ba &  0.13 & $-$0.18 & 0.00 & $-$0.12 & 0.25 \\
\hline\noalign{\vskip 3pt} 
\end{tabular}
}
\end{table}

\clearpage

\begin{figure}
 \begin{center}
  \includegraphics[width=15cm]{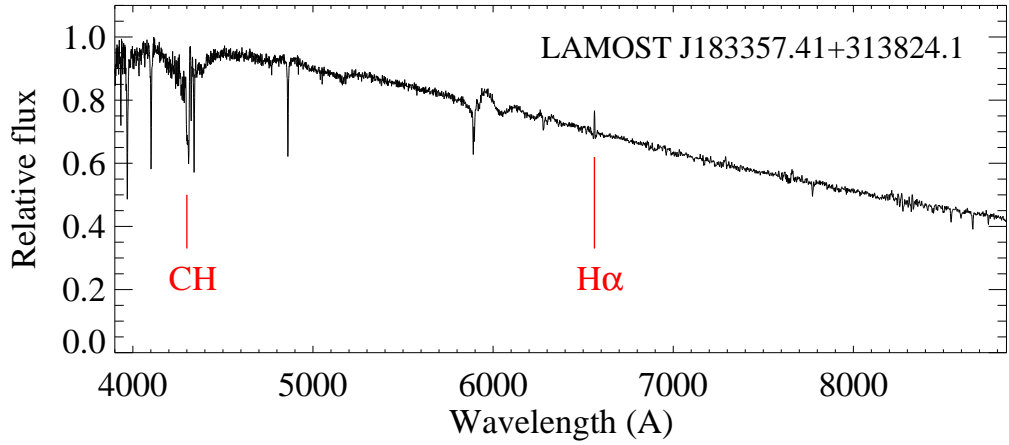}
 \end{center}
 \caption{The medium-resolution spectrum of CC Lyr (LAMOST J~183357.41+313824.1) obtained with LAMOST. The position of CH G-band and H$\alpha$ (emission) are presented.}\label{fig:lamost}
\end{figure}

\begin{figure}
 \begin{center}
  \includegraphics[width=7cm]{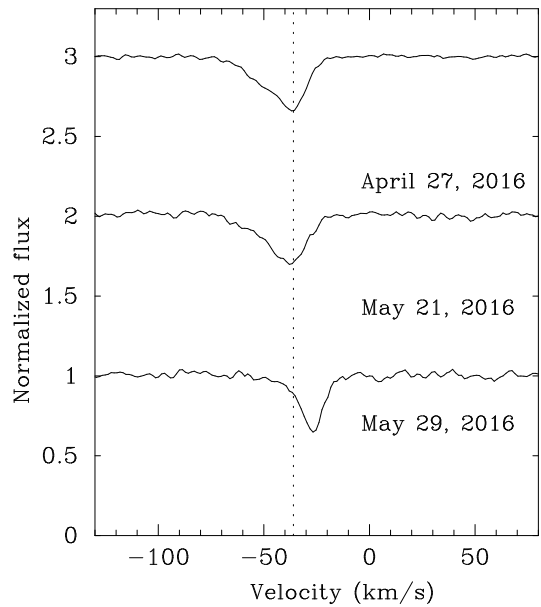}
 \end{center}
 \caption{The absorption feature of C I 5052~{\AA} line. The spectra
   of April 27 and May 21 are vertically shifted. The vertical dotted
   line is the radial velocity determined from other absorption lines
   for the April 27 spectrum.}\label{fig:c1}
\end{figure}

\begin{figure}
 \begin{center}
  \includegraphics[width=7cm]{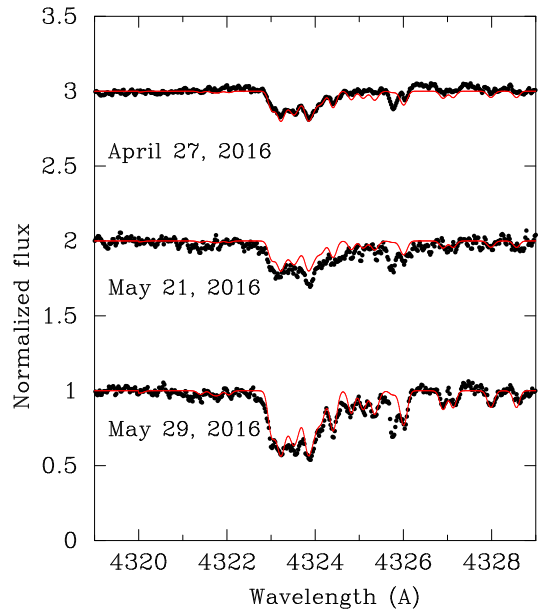}
 \end{center}
 \caption{The CH $A-X$ band of CC Lyr obtained at three epochs
   (dots). The spectra of April 27 and May 21 are vertically
   shifted. The red lines are synthetic spectra for $T_{\rm
     eff}=6000$~K for the spectra obtained on April 27 and May 21 and
   for $T_{\rm eff}=5750$~K for the May 29 spectrum.}\label{fig:ch}
\end{figure}

\begin{figure}
 \begin{center}
  \includegraphics[width=7cm]{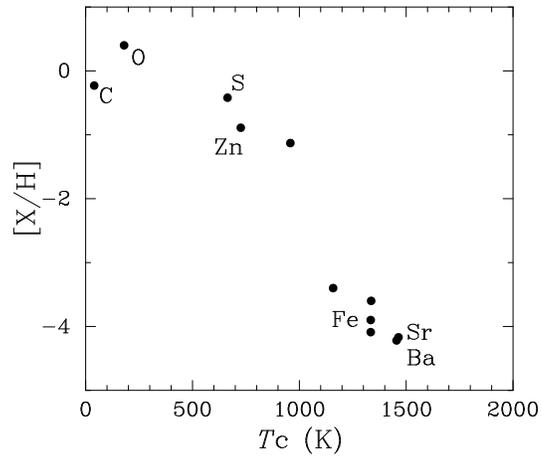}
 \end{center}
 \caption{Derived elemental abundance ratios ([X/H]) as a function of
   the condensation temperatures.}\label{fig:abtc}
\end{figure}

\begin{figure}
 \begin{center}
  \includegraphics[width=7cm]{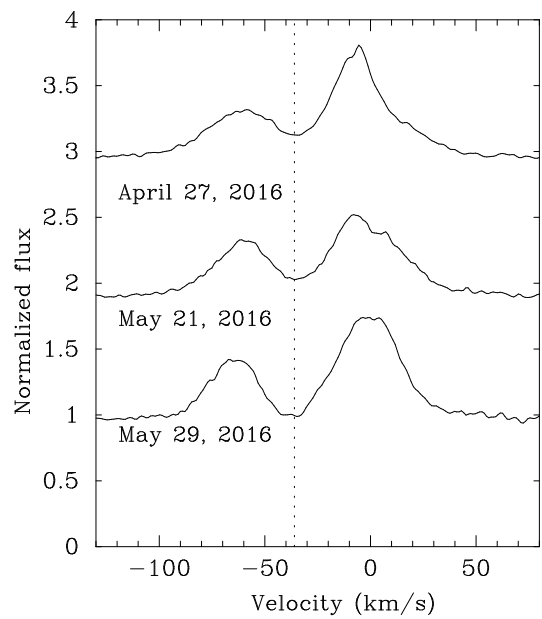}
 \end{center}
 \caption{The same as Figure~\ref{fig:c1}, but for the emission feature of H$\alpha$. }\label{fig:ha}
\end{figure}

\end{document}